# Performance Evaluation of Query Plan Recommendation with Apache Hadoop and Apache Spark

Elham Azhir [1,2], Mehdi Hosseinzadeh [3,4], Faheem Khan [5] and Amir Mosavi [6,*]

[1] Department of Computer Engineering, Science and Research Branch, Islamic Azad University, Tehran 1477893855, Iran; e.azhir@mci.ir
[2] Research and Development Center, Mobile Telecommunication Company of Iran, Tehran, Iran
[3] Mental Health Research Center, Psychosocial Health Research Institute, Iran University of Medical Sciences, Tehran 1449614535, Iran
[4] Computer Science, University of Human Development, Sulaymaniyah 0778-6, Iraq
[5] Department of Computer Engineering, Gachon University, Seongnam 13120, Korea
[6] Faculty of Civil Engineering, Technische Universität Dresden, 01069 Dresden, Germany
* Correspondence: faheem@gachon.ac.kr (F.K.); amir.mosavi@mailbox.tu-dresden.de (A.M.)

**Abstract:** Access plan recommendation is a query optimization approach that executes new queries using prior created query execution plans (QEPs). The query optimizer divides the query space into clusters in the mentioned method. However, traditional clustering algorithms take a significant amount of execution time for clustering such large datasets. The MapReduce distributed computing model provides efficient solutions for storing and processing vast quantities of data. Apache Spark and Apache Hadoop frameworks are used in the present investigation to cluster different sizes of query datasets in the MapReduce-based access plan recommendation method. The performance evaluation is performed based on execution time. The results of the experiments demonstrated the effectiveness of parallel query clustering in achieving high scalability. Furthermore, Apache Spark achieved better performance than Apache Hadoop, reaching an average speedup of 2x.

## 1. Introduction

A query's cost model includes local processing costs and communication costs between nodes in distributed query processing [1–3]. The search space of possible query execution plans (QEPs) can be fairly ample based on the complexity of the query. The query optimizer cannot efficiently search through all possible QEPs to find the most efficient QEP. The access plan recommendation method is introduced to decrease the query optimization costs by reusing the previously generated QEPs [4–6]. The access plan recommendation technique looks for textual similarities between the novel query and prior ones in order to reuse prior executed query plans for future ones. In the presented process, the query optimizer makes use of the likeness of query statements to execute new queries. However, clustering large query sets becomes a problem for traditional clustering algorithms due to the high processing time [7,8]. Various query plan prediction techniques are introduced [5,6] to identify the similarity between the queries using traditional clustering algorithms. For instance, Zahir, et al. [6] proposed an approach for recommending query plans that is efficient and effective. The SQL semantics are used for similarity detection. The expectation-maximization (EM) and K-means algorithms are applied in the proposed method. In addition, association rule (AR), naive Bayes (NB), and support vector machine (SVM) classification algorithms have been used in another study [5] for query plan prediction. The findings demonstrate that the AR algorithm is more accurate than the SVM and NB techniques in forecasting. It can be observed that the presented approaches reduce the expenses related to optimizing. Azhir, et al. [7] also presented a novel query plan recommendation technique depending on DBSCAN and NSGA-II algorithms for enhancing prediction accuracy. The outcomes related to the suggested method are compared to traditional K-means and DBSCAN. Based on the obtained outcomes, the presented approach achieves better performance in terms of accuracy. These papers aim to reuse previous query plans. However, clustering large query sets is one of the major challenges for such conventional clustering methods, due to the required processing time.

MapReduce is a distributed programming model for processing massive amounts of data [9,10]. MapReduce is an efficient method for clustering large query sets in a reasonable amount of time. Apache Hadoop and Apache Spark [11] are two of the most popular frameworks used for executing MapReduce programs. Apache Hadoop is a framework that is mainly used for scalable and distributed processing in the cloud [10,12]. Hadoop processes huge quantities of data by dividing tasks into smaller tasks [9]. It frequently writes and reads data from the disk. It can decrease the performance of the system. However, Apache Spark provides in-memory computations for reducing the number of the read/write cycle to disk [13]. We will go through a few clustering approaches that are built on parallel structures in the following sections. Elsayed, et al. [14] proposed a novel model to cluster intensive data documents. The WordNet ontology is used with bisecting K-means in order to utilize the semantic relations between words. The proposed model is based on distributed implementation for the bisecting K-means using the MapReduce programming model to improve document clustering results. The trial results have revealed that lexical categories for nouns enhanced internal evaluation measures of document clustering and decreased the document's features. According to the trial outcomes, the proposed approach obtained better results in text clustering than the typical K-means approach. Zewen and Yao [15] presented a parallel text clustering algorithm for the cloud environment. In this paper, the classical Jarvis–Patrick (JP) algorithm is implemented using the MapReduce programming model. In the proposed method, the shared nearest neighbor (SNN) table of the JP algorithm is created using MapReduce parallel processing. The results indicate the efficiency of the parallelization of the JP algorithm. Furthermore, high scalability is a key benefit of the technique. In addition, Li, et al. [16] suggested a new parallel method for clustering the text documents. The proposed algorithm is a parallel form of sequential K-means based on neighbors (SKBN). This parallel method, called PKBN, uses data parallelism in its three main phases, which are as follows: creating the neighbor matrix, choosing the primary centroids grounded on ranks, and repeating the loop of the assignment phase and the update phase. The neighbor matrix of the proposed algorithm is generated using a novel parallel pair-generating technique called PG-New. The results proved that the decrease in communication overheads is accorded to the PG-New. Furthermore, the scalability of PKBN is improved using the proposed parallelism. This paper aims to enhance the recommendation time of the access plan recommendation method. The term frequency (TF) method [17] and cosine measure with a feature representation of SQL query language are used in the presented access plan recommendation method. In the present article, a parallel MapReduce model is applied to sped up the query clustering operation in Apache Hadoop [18]. Furthermore, the performance of the presented access plan recommendation method [18] is improved using the implementation in Apache Spark, which is a in-memory distributed data processing engine. The following list underlines the article's key contributions:

(i) Presenting a parallel query plan recommendation method using the MapReduce model.
(ii) Implementing the parallel query plan recommendation technique using Apache Hadoop and Apache Spark distributed frameworks.
(iii) Evaluating the performance of the implemented algorithm with multiple-query datasets.

In Section 2, Apache Hadoop and Apache Spark distributed processing frameworks are presented. The parallel query plan recommendation method is introduced in Section 3. Our test environment is illustrated in Section 4. The findings of the simulation results are provided in Sections 5 and 6. Finally, Section 7 reviews the main points and recommends some suggestions for future research.

## 2. Overview and Background

In this part, two commonly used cloud frameworks are presented. Hadoop is a framework for storing and processing large quantities of data. It is based on the MapReduce programming model for parallel processing. However, Spark is used for real-time data analysis and can achieve higher performance than Hadoop MapReduce using the main memory.

*2.1 Apache Hadoop*

Apache Hadoop is an effective open-source platform for reliable, scalable, and distributed processing of large amounts of data. This framework can work with thousands of nodes and several petabytes of data. Generally, Hadoop has a set of modules for processing large quantities of distributed data on different nodes across the Hadoop cluster. Therefore, Hadoop is considered a collection that includes data gathering, storage, analysis, and maintenance services. There are two key parts in the core of Hadoop [19]: (1) the storage part is called the Hadoop Distributed File System (HDFS), which is responsible for partitioning, storing, and retrieving large files on a Hadoop cluster, and (2) the processing part is called MapReduce, which is responsible for analyzing and processing distributed data. Hadoop is based on master-worker architecture for data storage and distributed processing using

HDFS and MapReduce. In the Hadoop cluster, there are a master node and several worker nodes. In addition, a layered architecture is also used in Hadoop, with HDFS on the bottom layer and the MapReduce layer on it. Hadoop offers parallel and distributed processing on big data sets through the MapReduce model [10]. MapReduce is a programming model for parallel data processing that provides great scalability on a Hadoop cluster [9].

*2.2. MapReduce*

MapReduce distributes input data across different machines. In the MapReduce model, the processing operation is divided between several nodes. Therefore, MapReduce helps process big data through distribution. Data are processed in parallel by several machines instead of one machine. This means that the data processing time is greatly reduced. Figure 1 illustrates the MapReduce process. First, the client sends MapReduce execution requests to the JobTracker. JobTracker splits jobs into tasks among the cluster nodes. Furthermore, the JobTracker talks to the NameNode to find the DataNodes that contain the data to be used in processing. TaskTrackers perform processing independently and in parallel (Mapper).

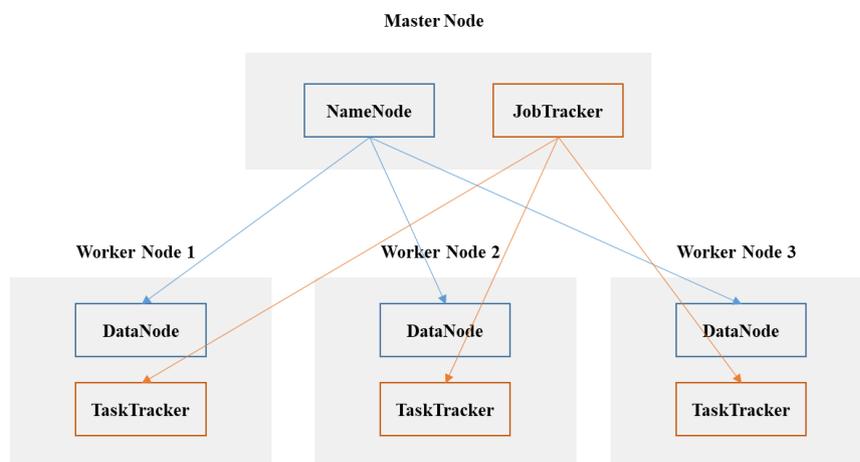

**Figure 1.** The MapReduce process in Hadoop.

The MapReduce processing model includes the following two key phases: the map phase and the reduce phase. As shown in Figure 2, the map phase takes key-value pairs as input and implements some processing on this input. The results of the map phase are passed to the reduce phase in the form of key/value pairs. Finally, the map phase's findings are processed in the reduce phase.

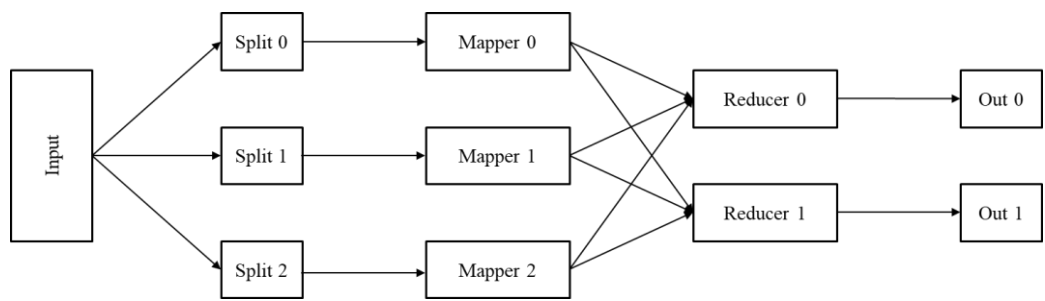

**Figure 2.** Hadoop MapReduce framework.

*2.3. Apache Spark*

Apache Spark is a framework for processing batch and stream data in a distributed computing manner. The Spark is introduced to increase processing speed and enhance complex processing. Spark uses in-memory computing to speed up data processing, which is more efficient than the Hadoop model. Spark speed is nearly 100 times faster than Hadoop when data are stored in memory and up to tens of times faster than when all the data are not in memory. Apache Spark combines various modules, such as MLib machine learning libraries, GraphX for large-scale graph analytics, Spark SQL for relational data analysis, and Spark Streaming for streaming data analysis [20]. Iterative algorithms, graph analytics, and interactive data analysis algorithms all perform better. Spark uses Hadoop YARN or Apache Mesos as a cluster manager and it has a distributed storage system. Spark supports a wide variety of distributed storage, including HDFS, Cassandra, OpenStack Swift, and Amazon S3. The resilient distributed dataset (RDD) and directed acyclic graph (DAG) execution engine are two key concepts in Spark [11]. RDD is the first distributed memory abstraction provided by Spark. It provides in-memory computation on large distributed clusters with high fault-tolerance. The Spark driver program creates RDD and divides it among different nodes. Transformations and actions are two kinds of operations executed by RDD. Transformations generated new datasets from the input RDD (map), and actions return a value after performing calculations on the dataset (reduce). Spark can cache RDDs on memory and repeat the MapReduce operation without performance overhead. Therefore, Spark has a performance advantage for recursive algorithms. When an action is executed on an RDD by a user, a DAG is created according to the transformation dependencies. This enhances the performance and eliminates the multistage execution model of MapReduce.

## 3. Methodology

The access plan recommendation method uses earlier query plans for new queries. Initially, optimal QEPs are generated for a query group in the system log. The output of the query optimizer is a set of optimal QEPs of the existing queries. This technique transforms query statements into feature vectors, which are then compared to verify how similar they are. Therefore, when a new query is created, the query vector is produced using the query statement. Then, the similarity degree between the novel query and the other ones placed in the clusters is calculated, and the ideal query plan is extracted. If the query is not similar to the existing clusters, the optimizer generates the optimal query plan to update the clustering. Query preprocessing, tokenizing, feature weighting based on TF, similarity measurement, and plan recommendation are the five key stages of the access plan recommendation method. Accordingly, various stages are presented in Figure 3, and the way the presented method is converted to reduce/map flows is explained in the following section.

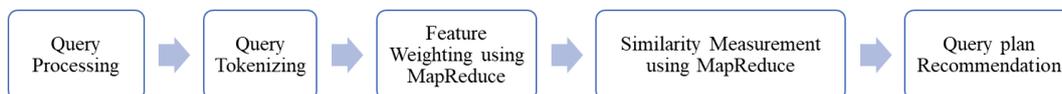

**Figure 3.** The map-reduce-based access plan recommendation.

### 3.1. Query Preprocessing

In the presented method, the SQL statements are manipulated using the JSQL parser. In order to enhance the clustering quality, a custom JSQL parser library (http://jsqlparser.sourceforge.net/, accessed on 23 February 2020) and different normalization rules to increase the clustering quality are applied to queries' expressions [7]. The JSQL parser rewrites the queries' text to eliminate string constants, table aliases, column aliases, syntax sugar, and namespaces [21].

## 3.2. Query Tokenizing

The clustering algorithms divide data into different groups using a feature space. In the tokenization, the queries are represented in a feature space. Therefore, the tokenization process split the queries' texts into words. The words' frequencies are also calculated as the queries' features for clustering in the next part. In the tokenizing stage, the parser breaks the query text into tokens. In addition, the parser qualifies tokens by utilizing the SQL clauses, including FROM, SELECT, GROUP BY, WHERE, and ORDER BY (Table 1).

**Table 1.** An example of SQL query tokenization.

| Raw SQL Query | Tokenized SQL |
|---|---|
| **SELECT** name | **SELECT** name |
| **FROM** instructor | **FROM** instructor |
| **WHERE** salary > 90,000 **AND** salary < 100,000 | **WHERE** salary |

## 3.3. Feature Weighting based on TF using MapReduce

The frequency of a word in a document is calculated using the TF method. In this paper, the TF method assesses the importance of a feature in a query. In the presented access plan recommendation method, the created tokens (features) are weighted by frequency. The number of times the feature f appears in query q is shown as tf (f,q) [17]. In this stage, the term frequency algorithm computes the weight of the word in the query vector using the frequency of each word [22]. Table 2 describes an example of the TF calculation for a simple query. As shown in Tables 1–2, to calculate the weight of the term "WHERE salary," it appears twice in the query [17].

**Table 2.** An example of a feature vector.

| Tokenized SQL | Feature Vector |
|---|---|
| **SELECT** name | {'SELECT name' →1, |
| **FROM** instructor | 'FROM instructor →1', |
| **WHERE** salary | 'WHERE salary' →2}, |

The parallelized feature weighting algorithm in MapReduce is presented in Algorithms 1–3. First, Algorithm 1 counts the number of each term in each query. Algorithm 2 calculates the whole terms of each query. Then, Algorithm 3 calculates the weights of each feature by using the TF method.

| **Algorithm 1:** The number of each term in every query |
|---|
| **Input:** query |
| **Output:** The total number of occurrences of each token (term) in every query (queryID); ((term, queryID), sum) |
| 1: **class** Mapper: A key/value pair is generated for each query. |
| 2: **function** map (query) |
| 3: **for all** term ∈ query.split() **do**//The query is divided into smaller units (term) |
| 4: emit ((term, queryID), 1)   //The number 1 is assigend as the value to each term. |
| 5: **end for** |
| 6: **end function** |
| 7: **end class** |
| 1: **class** Reducer: The total number of occurrences of each term in a query is counted. |
| 2: **function** reduce ((term, queryID), counts [ $c_1$, $c_2$, …]) |
| 3: sum = 0 |
| 4: **for all** count ∈ counts **do** |
| 5: sum += count |

6: **end for**
7: emit ((term, queryID), sum)     //Key: term and queryID, Value: The number of repetitions of a term in a query.
8: **end function**
9: **end class**

---

**Algorithm 2:** The total terms of each query

**Input:** The number of each token in each query ((term, queryID), sum)
**Output:** The total number of terms (sum) in a query ((queryID, N), (term, sum))
1: **class** Mapper: All terms in each query are grouped.
2: **function** map ((term, queryID), sum)
3: **for all** element ∈ (term, queryID) **do**
4: emit (queryID, (term, sum))     // Key: queryID, Value: The term and the number of repetitions of each term in a query.
5: **end for**
6: **end function**
7: **end class**

1: **class** Reducer: The total number of terms in a query is computed.
2: **function** reduce (queryID, (term, sum))
3: N = 0                                              //N is the total number of terms in a query.
4: **for all** tuple ∈ (term, sum) **do**
5: N = N + sum
6: **end for**
7: emit ((queryID, N), (term, sum))
8: **end function**
9: **end class**

---

**Algorithm 3:** The weights of each feature

**Input:** The total number of terms in a query ((queryID, N), (term, sum))
**Output:** The weights of each feature ((queryID), (term, tf))
1: **class** Mapper:
2: **function** map (queryID, N), (term, sum))
3: **for all** element ∈ (term, sum) **do**
4: emit ((term, (queryID, sum, N))
5: **end for**
6: **end function**
7: **end class**

1: **class** Reducer: The term frequency is calculated.
2: **function** reduce ((term, (queryID, sum, N))
3: **for all** element in (queryID, sum, N) **do**
5: tf = sum/N;
6: **end for**
7: emit ((queryID), (term, tf));
8: **end function**
9: **end class**

*3.4. Similarity Measurement using MapReduce*

In the presented method, the Cosine similarity [23] is applied to find similar queries. To measure the similarity between the two query vectors $\vec{q_1}$ and $\vec{q_2}$, the cosine of the angle between $\vec{q_1}$ and $\vec{q_2}$ is calculated using (2) [23].

**Algorithm 4:** The Cosine similarity of the produced query vectors
**Input:** query vectors
**Output:** The Cosine similarity between the produced query vectors
1: **class** Mapper: The key/value pair is formed.
2: **function** map (queries)
3: **for** i = 0 to n                                    //Where n is the total number of queries.
4:   **for** j = i + 1 to n
5:     emit ((q [i].id, q[j].id), (q[i].tf, q[j].tf)) //The term frequency values are assigned as the value.
6:   **end for**
7: **end for**
8: **end function**
9: **end class**
1: **class** Reducer: The Cosine similarity is calculated for each pair of queries.
2: **function** reduce ((q$_1$, q$_2$), (q$_1$.tf, q$_2$.tf))
3: $cosine = sum\ (q_1.\text{tf} \times q_2.\text{tf})/(sqrt\ (sum\ (q_1.tf)^2) \times sqrt\ (sum\ (q_2.tf)^2)$
4: emit ((q$_1$, q$_2$), cosine)
5: **end function**
6: **end class**

$$cos(q_1, q_2) = \frac{\vec{q_1} \cdot \vec{q_2}}{|\vec{q_1}| \times |\vec{q_2}|} = \frac{\sum_{i=1}^{t} w_{i,1} \times w_{i,2}}{\sqrt{\sum_{i=1}^{t} w^2_{i,1}} \times \sqrt{\sum_{i=1}^{t} w^2_{i,2}}}$$

where the weight of the words in each query vector is divided by the length of the words in that query. Algorithm 4 may be used to compute the Cosine similarity between the produced query vectors in parallel [22].

*3.5. Plan Recommendation*

In this part, the recommendation process is performed; for a new query, the query is assigned to one cluster according to its detected similarity. The recommendation is made using the identified similarities between the novel query and the old ones whose access plans are accessible in the memory. If a similarity is identified, the recommender system suggests that the query optimizer reuses the old query execution plan to execute the new incoming query. DBSCAN is a clustering method based on density that can find clusters of any shape. In the introduced recommender system, DBSCAN clustering is performed from the precomputed distance matrix. A few main definitions of DBSCAN are [24] as follows:

**Definition 1.** *The $Eps$ neighborhood of a point $p$ is a set of points within distance $Eps$ from point $p$. It can be expressed as $N_{Eps}(p) = \{q \in D\ |dis(p, q) \ll Eps\}$.*

**Definition 2.** *$p$ is directly density-reachable from the point $q$ if $p$ is in the $Eps$ radius of $q$ (i.e., $p \in N_{Eps}(q)$ ) and $q$ is a center object (i.e., $|N_{Eps}(q)| \gg Minpts$).*

**Definition 3.** *$p$ is density reachable from the point q if there is a directly density reachable connection between two points.*

The DBSCAN clustering technique starts by choosing a random point $p$ and then retrieving density-reachable points from point $p$. The algorithm creates a cluster if $p$ is a core object. If no point can be reached using density from point $p$, the algorithm moves on to the subsequent point in the given dataset. This technique is carried out for each of the points. To divide queries into similar clusters, the R DBSCAN package is used [25].

## 4. Experimental SetUp

Several experiments are conducted to evaluate the effectiveness of the presented parallel access plan recommender system. The employed set-up for experiments is presented in this part. Virtualbox 6.1.16 was used to create the virtual environment in a physical system, with 16 GB of RAM and a 2.8 GHz Intel Core i7 processor. The tests were carried out on a Hadoop cluster with varying numbers of virtual nodes. One node is the master, while the others are workers. On the master and worker nodes, the Red Hat (64bit) Linux Operating System (OS) is installed. The master node has four virtual cores and eight gigabytes of memory. Each worker node has two virtual cores and three gigabytes of RAM. In addition, Hadoop 2.8.0 was installed. The IIT Bombay dataset [26] is also used to create simple selection query sets. Therefore, the bind variables of the queries are substituted with different data values. Table 3 contains three query sets. A distinct hash value is assigned to a query's QEP when it is run in Oracle. Each query set is manually classified based on the distinctive value of its QEPs. The S2 query set, for instance, has 593 queries divided into 14 classes.

Table 3. Features of the generated query sets.

| Features | Dataset | Number of Classes | Number of Individuals |
|---|---|---|---|
| Selection | S1 | 7 | 235 |
|  | S2 | 14 | 593 |
|  | S3 | 18 | 1150 |

## 5. Results

The performance of the presented parallel access plan recommendation method is evaluated using both Hadoop [27] and Spark frameworks in terms of execution time. The presented algorithm is evaluated on the three created query sets described in Table 3. Table 5 demonstrated the presented technique's implementation time for various numbers of queries with diverse worker nodes. As shown in Table 4, Hadoop took a long time on the data sets with large query sizes. However, it performs better when the number of queries is reduced. Similar results were also obtained from Spark. In Spark's case, the results were not significantly different when we reduced the number of queries.

Table 4. Run-time (Seconds) of Hadoop and Spark with the change in number of queries.

| Number of Worker Nodes | Number of Queries | Hadoop | Spark |
|---|---|---|---|
| 1 | 235 | 59 | 39 |
|  | 593 | 152 | 58 |
|  | 1150 | 260 | 105 |
| 3 | 235 | 57 | 25 |
|  | 593 | 150 | 43 |
|  | 1150 | 255 | 92 |
| 5 | 235 | 51 | 19 |
|  | 593 | 130 | 33 |
|  | 1150 | 224 | 71 |
| 7 | 235 | 45 | 15 |
|  | 593 | 119 | 25 |
|  | 1150 | 197 | 50 |
| 9 | 235 | 45 | 11 |
|  | 593 | 119 | 22 |
|  | 1150 | 190 | 42 |

Because of Hadoops' inherent nature of using disk I/O for its computation, which increases Hadoop execution time as the data size increases, it is more beneficial to use Spark instead of Hadoop for large datasets. Figure 4 shows that multi-node cluster implementation always outperforms the stand-alone implementation. Therefore the MapReduce model performs better in the distributed mode rather than on a single machine. Furthermore, it is obvious that memory-based Spark performs better than disk-based MapReduce for all query sets. The outcomes revealed that the speedup value is enhanced along with the number of nodes in Hadoop and Spark. However, Apache Spark scales performed better than Hadoop when the number of nodes was enhanced.

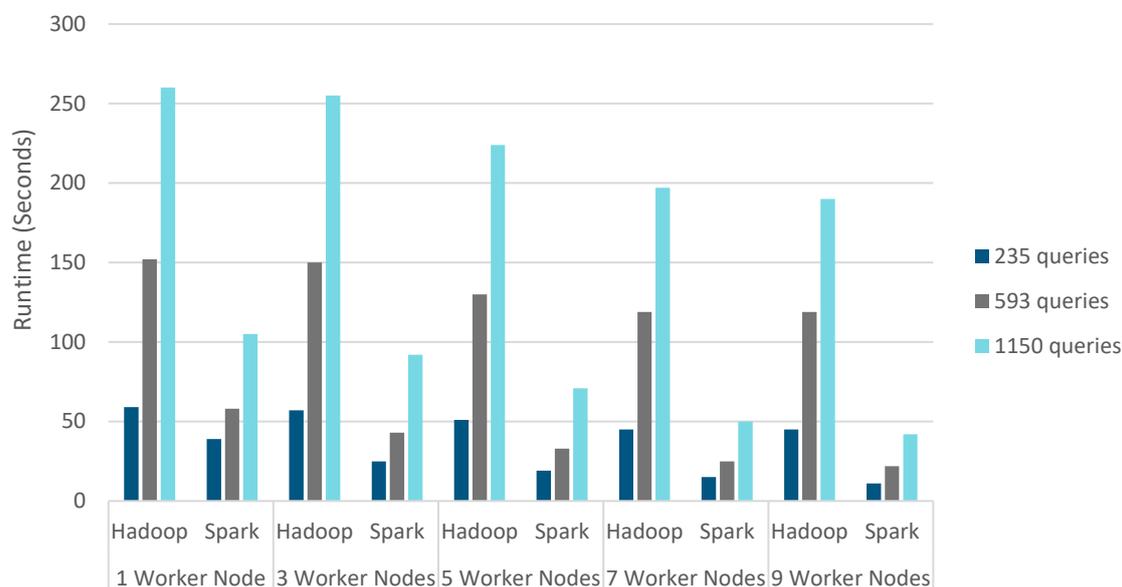

**Figure 4.** Run-time of Hadoop on different nodes.

The query optimization problem in large-scale and distributed databases is NP-hard in nature. The query optimizer's complexity increases with the complexity of the query. Therefore, the large space of alternative query plans should be explored by the query optimizer to generate optimal QEPs. Recommendation-based methods help the query optimizer to recognize similarities between old and future queries. An efficient parallel processing method based on MapReduce is used in this paper to increase the efficiency of the proposed access plan recommendation method [7]. The presented MapReduce algorithm is executed on different query sets. The trial outcomes have shown that the presented parallel recommendation method can process large query sets with time-efficiency. As shown in Figure 4, as the number of queries increases, the efficiency of parallel computing will be higher. The results have shown that the speedup increases with the number of queries. It was also concluded that the proposed method in the Spark platform could more effectively solve the large-scale query set processing with reasonable processing time.

## 6. Conclusions

This paper aimed to enhance the recommendation time in the access plan recommendation method. For this purpose, the access plan recommendation method was implemented with two common cloud-based frameworks, Apache Hadoop and Apache Spark. First, the query expression was standardized, and query vectors were generated. Then, parallelized feature weighting and similarity measurement algorithms were designed to create the queries' weight matrix. At last, the recommendation was made using the DBSCAN clustering algorithm. The performances of Hadoop and Spark were compared by investigating the execution time. Several experiments were performed with different cluster nodes and query sets. The experiments showed that Spark outperforms Hadoop in all cases due to its efficient utilization of main memory. However, Apache Hadoop was more cost-effective. Future work is recommended to investigate the performance of other clustering and classification algorithms for query plan recommendations regarding execution time and accuracy. Furthermore, additional tests will be beneficial in evaluating the suggested strategy's performance for more complicated query sets. Finally, the optimization of the MapReduce process using the partitioning and indexing techniques can be addressed by future works.

**Author's Contributions:** Conceptualization, E.A. and M.H.; Methodology, E.A. and M.H.; Software, E.A.; Validation, E.A., F.K. and A.M.; Resources, E.A.; Data Curation, E.A. and A.M.; Writing—Original Draft Preparation, E.A.; Writing—Review and Editing, E.A. and F.K.; Visualization, E.A.; Supervision, A.M.; Project Administration, F.K. All authors have read and agreed to the published version of the manuscript.

**Funding:** This research received no external funding.

**Institutional Review Board Statement:** Not applicable.

**Informed Consent Statement:** Not applicable.

**Data Availability Statement:** Not applicable.

**Conflicts of Interest:** The authors declare no conflict of interest.